\documentclass[aps,prb,reprint,superscriptaddress,showpacs,draft]{revtex4-1}
\usepackage{amssymb,amsmath,graphicx}

\begin{document}

\title{Approximate diagonalization method for many-fermion Hamiltonians}

\author{Jonathan E. Moussa}
\email[]{godotalgorithm@gmail.com}
\affiliation{Center for Computational Materials, Institute for Computational Engineering and Science, University of Texas, Austin, Texas 78712, USA.}

\date{\today}

\begin{abstract}
The limits of direct unitary transformation of many-fermion Hamiltonians are explored.
Practical application of such transformations requires that effective
 many-body interactions be discarded over the course of a calculation.
The truncation of the Hamiltonian leads to finite errors and in some cases divergences.
A new formalism is proposed to manage errors and avoid divergences.
Removing all interactions from a many-fermion Hamiltonian reduces it to fermion number operators
 allowing for direct calculation of eigenvalues.
If the same transformations are applied to the bare fermions, eigenfermions are produced whose Slater determinants form eigenstates.
This enables a hierarchy of diagonalization methods of increasing accuracy as fewer interactions are discarded from the Hamiltonian.
\end{abstract}


\maketitle

\section{Introduction}

Simulation of interacting fermions is difficult.
The root of the difficulty is the large size of the fermion configuration space,
 which is exponential in the number of fermion degrees of freedom.
Brute force calculations on the large space are only tractable
 for small model systems and small molecules \cite{Olsen.FCI}.
Approximations are necessary, but there is no concensus on an approximate method
 that is sufficiently accurate and efficient for systems of interest.
Development is split mainly between a few popular approaches
 based on well-established ideas: many-body perturbation theory
 with a variety of resummations and empirical parameterization \cite{Onida.TDDFT.GWBSE},
 quantum Monte Carlo to statistically sample large configuration spaces \cite{Foulkes.QMC},
 and methods based on renormalization group (RG) principles.
The two most popular RG methods are based on the extreme limits of
 one \cite{Schollwock.DMRG} and infinite \cite{Georges.DMFT} spatial dimensions.

Another, less popular RG approach exists that is based on preserving the form
 of a many-fermion Hamiltonian under unitary transformations \cite{Wegner.flow}.
With no reference to spatial dimension or scale,
 it is less fundamentally restricted than other RG approaches.
Its primary use to date has been in decoupling weakly correlated fermions from a system
 to produce a smaller strongly correlated subsystem to be solved using other methods
 \cite{White.canonical.diagonalization,Yanai.canonical.transformation}.
Attempts to decouple all fermions with this method have resulted in 
 slow convergence and the appearance of numerical divergences \cite{White.canonical.diagonalization,Wegner.oldflow}.
What remains unclear are the source of these problems, and whether they
 present a fundamental barrier to improvement or merely a technical barrier.

The work presented in this paper addresses the technical problems of
 previous many-fermion transformation methods.
The fundamental source of error in these methods is the truncation of the Hamiltonian
 after each transformation, to remove terms outside a prescribed Hamiltonian form.
Divergences resulting from truncation can be eliminated by conserving a set
 of quantities that are naturally conserved by exact unitary transformation.
The restrictions placed on operator truncation specify a unique form.
In a flow equation framework, the continous transformation of the Hamiltonian
 results in continuous growth of truncation errors.
All previous errors get locked into the solution.
To minimize the total truncation error, the continuous transformation
 is grouped into discrete fragments, each of which is carefully optimized
 based on an error minimization criteria.
Some finite amount of truncation error inevitably remains and can only be reduced
 further by truncating fewer terms from the transformed Hamiltonian.

Completely decoupling all fermions in a many-fermion Hamilonian
 reduces it to a diagonal form containing only fermion number operators.
The transformation that diagonalizes a many-fermion Hamiltonian can be
 applied either directly to the Hamiltonian or the elementary fermion operators.
The transformed fermions are called \textit{eigenfermions}.
Eigenstates are Slater determinants of eigenfermions.
Whereas the eigenvalue decomposition of a matrix produces a list
 of all eigenvalues, the \textit{eigenfermion decomposition}
 of a many-fermion Hamiltonian produces a function of eigenfermion number operators.
Eigenvalues are evaluated by replacing number operators with occupation numbers.
The complete set of eigenstates is parameterized by a configuration space
 of eigenfermion occupation numbers much like the states of a classical system
 lie in a configuration space of classical variables.
Based on this analogy, the diagonalization of a many-fermion Hamiltonian
 can be interpretted as a quantum-to-classical mapping.

The paper proceeds as follows.
Section \ref{eigen_section} defines and discusses the concept of an eigenfermion.
Section \ref{eigen_theory} constructs a general mathematical theory of truncated unitary transformations
 and truncated eigenvalue decomposition.
Section \ref{for_fermions} applies the general theory to the many-fermion case.
Section \ref{TED_framework} discusses what can be computed as a result of a
 \textit{truncated eigenfermion decomposition} and the associated computational costs.

\section{The eigenfermion concept \label{eigen_section}}

The many-fermion Hamiltonians that describe physical systems typically
 contain only 1\&2-fermion interactions.
The structure of these Hamiltonians is compactly encoded
 in the standard language of second quantization,
\begin{equation} \label{hamiltonian}
 \hat{H} = \sum_{i,j} h_{ij} \hat{c}_i^\dag \hat{c}_j + \sum_{i,j,k,l} V_{ijkl} \hat{c}_i^\dag \hat{c}_j^\dag \hat{c}_k \hat{c}_l.
\end{equation}
If $\hat{H}$ is diagonalized in the basis of Slater determinants by a
 unitary transformation $\hat{U}$, then the transformed Hamiltonian
 contains only fermion number operators, $\hat{n}_i = \hat{c}_i^\dag \hat{c}_i$,
\begin{equation} \label{diagH}
 \hat{U}^\dag \hat{H} \hat{U} = E_0 + \sum_{i} E_i \hat{n}_i + \sum_{i,j} E_{ij} \hat{n}_i \hat{n}_j + \cdots .
\end{equation}
All possible products of $\hat{n}_i$ operators can appear in this expression.
The 1\&2-fermion form of $\hat{H}$ in Eq. (\ref{hamiltonian}) cannot generally be preserved
 by the diagonalization procedure.
The eigenvalues and eigenstates of $\hat{H}$ are parameterized
 by an occupation vector $\mathbf{f}$ with entries $f_i \in \{0,1\}$,
\begin{subequations} \label{eigenstuff} \begin{align} \label{eigenvalue}
 E(\mathbf{f}) &= E_0 + \sum_{i} E_i f_i + \sum_{i,j} E_{ij} f_i f_j + \cdots \\
 |\Psi(\mathbf{f})\rangle &= \hat{U}|\mathbf{f}\rangle = \hat{U} \prod_{i} [(1-f_i) + f_i \hat{c}_i^\dag] |0\rangle.
 \label{eigenvector} \end{align} \end{subequations}
Each $|\mathbf{f}\rangle$ is a Slater determinant and
 $|\mathbf{0}\rangle$ is the zero-fermion vacuum state.

Occupation vectors are convenient mathematical labels for eigenstates in Eq. (\ref{eigenstuff}),
 but to have physical significance they must describe the occupations of a physical object.
This object is a fermion and because of its special
 relationship to eigenstates, it is named an eigenfermion.
The $\hat{U}$ that diagonalizes $\hat{H}$ can be used to define
 eigenfermion operators from the bare fermions of the system,
\begin{equation} \label{qp_transform}
 \hat{q}_i = \hat{U} \hat{c}_i \hat{U}^\dag \ \ \mathrm{and} \ \ \hat{m}_i = \hat{q}_i^\dag \hat{q}_i.
\end{equation}
The anti-commutation relations of $\hat{c}_i$ are inherited by $\hat{q}_i$.
By rearranging Eq. (\ref{diagH}), the original Hamiltonian can be
 written solely in terms of eigenfermion number operators,
\begin{equation} \label{qp_ham}
 \hat{H} = E_0 + \sum_{i} E_i \hat{m}_i + \sum_{i,j} E_{ij} \hat{m}_i \hat{m}_j + \cdots .
\end{equation}
The eigenstates can be written as a Slater determinant of eigenfermions
 acting on a new vacuum, $|\emptyset\rangle = \hat{U}|\mathbf{0}\rangle$,
\begin{equation} \label{qp_vec}
  |\Psi(\mathbf{f})\rangle = \prod_i [(1-f_i) + f_i \hat{q}_i^\dag] |\emptyset\rangle.
\end{equation}
If $\hat{U}$ preserves total fermion number, then $|\emptyset\rangle = |\mathbf{0}\rangle$.
It is clear from Eqs. (\ref{qp_ham}) and (\ref{qp_vec}) that occupation vectors labelling
 the eigenstates specify eigenfermion occupation numbers.

The only properties that eigenfermions are guaranteed to share
 with the bare fermions of a system are those preserved by symmetry.
Such symmetries must be explicitly preserved by $\hat{U}$ in Eq. (\ref{qp_transform}).
If total fermion number is conserved, then the number of eigenfermions in a state
 corresponds to the number of fermions.
If total fermion spin is conserved, then eigenfermions will have
 the same well-defined spin as the bare fermions.
If translational invariance is conserved,
 then eigenfermions will have a well defined crystal momentum.
Unless completely constrained by symmetry,
 the eigenfermions that diagonalize a Hamiltonian are non-unique.
However, it is possible to define a unique set of eigenfermions that are
 in some sense most similar to the bare fermions.

Practical calculations of Eq. (\ref{diagH}) will require approximations.
The generic form of an approximately diagonalized Hamiltonian is
\begin{equation} \label{approx_diag}
 \hat{U}^\dag \hat{H} \hat{U} = \hat{D} + \hat{R},
\end{equation}
 where $\hat{D}$ contains only fermion number operators
 and $\hat{R}$ is a residual interaction.
The physical effect of $\hat{R}$ is to scatter eigenfermions,
 reducing $|\Psi(\mathbf{f})\rangle$
 from an eigenstate to a finite-lifetime nearly-stationary state.
This situation resembles a theory of fermion quasiparticles
 that are adiabatically connected to the bare fermions.
Such a quasiparticle theory can be cast in the form of Eq. (\ref{approx_diag})
 with a 1-fermion diagonal term, $\hat{D} = \sum_i \epsilon_i \hat{n}_i$,
 and $\hat{R}$ that weakly scatters between a ground state $|\Psi(\mathbf{f}_{GS})\rangle$
 and certain few-quasiparticle excited states $|\Psi(\mathbf{f}_{X})\rangle$,
 $\langle \mathbf{f}_X | \hat{R} |\mathbf{f}_{GS} \rangle \approx 0$.
Eigenfermions generalize these fermion quasiparticles to
 allow for strong non-scattering interactions in $\hat{D}$ while
 requiring a weak residual interaction between all states, $\hat{R} \approx 0$.
If $\hat{R}$ is sufficiently small,
 then $\hat{U}$ exactly diagonalizes a perturbed Hamiltonian,
\begin{equation} \label{dirty_diag}
 \hat{U}^\dag (\hat{H} + \Delta \hat{H}) \hat{U} = \hat{D} \ \ \mathrm{with} \ \ \Delta \hat{H} = -\hat{U} \hat{R} \hat{U}^\dag.
\end{equation}
Constructing the exact solution to a system slightly different
 from what was intended is similar to experimenting on impure samples.

There are many distinct approaches in the literature
 \cite{Taube.UCC,Wegner.flow,Hubsch.projector.renormalization,White.canonical.diagonalization,Yanai.canonical.transformation,Mazziotti.ACSE}
 for approximating $\hat{U}^\dag \hat{H} \hat{U}$ for a
 general 1\&2-fermion Hamiltonian and a wide class of transformations.
Unfortunately, none of these methods have claimed to generally and reliably
 solve Eq. (\ref{approx_diag}) with a small residual error $\hat{R}$.
The deficiency is often attributed to nearly degenerate states or strong fermion correlations.
Inadequate closure of equations can result in uncontrolled growth of $\hat{R}$
 and numerical divergences \cite{White.canonical.diagonalization}.
No analysis has yet isolated the fundamental source of divergences.
No \textit{a priori} criteria has yet to guarantee the existence of solutions.
Recent progress has focused on unitary transformations that only partially diagonalize
 a Hamiltonian, leaving the rest of the problem to other many-fermion methods.
A lack of progress in full diagonalization might be blamed
 on the complexity of many-fermion algebra.
It proves productive to step back from the many-fermion case
 and construct a general theory of operator diagonalization
 in the presence of non-negligible truncation errors.

\section{Truncated eigenvalue decomposition theory \label{eigen_theory}}

The goal of this section is to develop a basic theory for the
 effect of truncation errors on continuous unitary transformations.
A continuous unitary transformation of an operator $\hat{X}$
 is defined by a differential equation of a real variable $\lambda$,
\begin{equation} \label{flow_ev}
 \frac{d}{d\lambda} \hat{X}(\lambda) = [\hat{A}(\lambda),\hat{X}(\lambda)],
\end{equation}
 specified by an anti-Hermitian generator function $\hat{A}(\lambda)$
 and starting from $\hat{X}(0) = \hat{X}$.
Operator truncation will modify this equation.
Diagonalization of a Hermitian operator $\hat{H}$ is possible if
 the modified form of Eq. (\ref{flow_ev}) can be evolved
 to a diagonalized form, $\hat{H}(\infty) = \hat{D}$.
A theory should provide \textit{a priori} conditions
 that guarantee the success of the diagonalization process.
Specifically, the evolution of $\hat{H}(\lambda)$ should contain
 no divergences and no stable, non-diagonal fixed points.

The form of truncation assumed by the theory is that $\hat{X}(\lambda)$
 is restricted to a given \textit{model subspace} of the full operator space.
There is a corresponding restriction of $\hat{A}(\lambda)$
 to a given \textit{generator subspace}.
If the elements of the model and generator subspaces form a Lie algebra,
 then no truncations are necessary and an established
 diagonalization method exists \cite{Wildberger.lie_diag}.
Otherwise, Eq. (\ref{flow_ev}) cannot be calculated exactly
 and a truncation procedure must be defined to close the equation.
For the application to many-fermion systems, a wide variety of
 physically motivated truncation procedures have been proposed.
What is more suitable for a general theory is a truncation that
 preserves $\mathrm{Tr}[\hat{X}^\dag(\lambda) \hat{X}(\lambda)]$
 as a conserved quantity over $\lambda$.
This criteria prevents divergences during the truncated transformation process
 and heavily constrains the form of truncation.

The constraint on truncation is clarified by
 elevating the space of operators to a Hilbert space,
 which requires the definition of an operator inner product,
\begin{equation}
 \langle \hat{X} ,\hat{Y} \rangle = \mathrm{Tr}[\hat{X}^\dag \hat{Y}].
\end{equation}
An operator inner product naturally defines an operator outer product $\hat{X} \otimes \hat{Y}$,
\begin{equation}
 (\hat{X} \otimes \hat{Y}) \hat{Z} = \hat{X} \langle \hat{Y},\hat{Z} \rangle,
\end{equation}
 which can be used to construct linear maps between operators,
 commonly referred to as superoperators.
Without constraint, the most general possible form for truncation of Eq. (\ref{flow_ev}) is
\begin{equation} \label{trunc_flow}
 \frac{d}{d\lambda} \hat{X}_M(\lambda) = \widehat{\mathcal{T}}(\lambda) [\hat{A}(\lambda),\hat{X}_M(\lambda)],
\end{equation}
 where $\widehat{\mathcal{T}}(\lambda)$ is a truncation superoperator
 acting on the result of the commutator.
$\hat{X}_M(\lambda)$ is an approximation of $\hat{X}(\lambda)$ restricted to the model subspace,
 intialized to $\hat{X}_M(0) = \widehat{\mathcal{T}}(0) \hat{X}$.
The range of $\widehat{\mathcal{T}}(\lambda)$ must be
 the model operator subspace to close the equation.

The target conserved quantity is now the
 natural norm of the operator Hilbert space,
\begin{equation}
 \|\hat{X}\| = \sqrt{\langle \hat{X},\hat{X}\rangle} = \sqrt{\mathrm{Tr(\hat{X}^\dag \hat{X})}}.
\end{equation}
If Eq. (\ref{trunc_flow}) is suggestively rewritten as
\begin{equation}
 \frac{d}{d\lambda} \hat{X}_M(\lambda) = \widehat{\mathcal{A}}(\lambda) \hat{X}_M(\lambda),
\end{equation}
 then $\widehat{\mathcal{A}}(\lambda)$ must be anti-Hermitian
 to conserve $\|\hat{X}_M(\lambda)\|$,
\begin{equation}
 \frac{d}{d\lambda} \|\hat{X}_M(\lambda)\|^2 = \langle \hat{X}_M(\lambda),[\widehat{\mathcal{A}}(\lambda)+\widehat{\mathcal{A}}^\dag(\lambda)]\hat{X}_M(\lambda)\rangle.
\end{equation}
Since $\widehat{\mathcal{T}}(\lambda)$ restricts the range of $\widehat{\mathcal{A}}(\lambda)$,
 anti-Hermicity correspondingly restricts the domain.

Specifying the constrained forms of $\widehat{\mathcal{A}}(\lambda)$ and $\widehat{\mathcal{T}}(\lambda)$
 requires the definition of a few more pieces of notation.
First, the commutator can be written as a superoperator,
 which is anti-Hermitian for an anti-Hermitian argument,
\begin{equation}
 \widehat{\mathcal{C}}[\hat{X}] \hat{Y} = [\hat{X},\hat{Y}].
\end{equation}
Second, a Hermitian projection superoperator is defined
 for the model subspace,
\begin{equation}
 \widehat{\mathcal{P}}_M = \sum_i \hat{M}_i \otimes \hat{M}_i,
\end{equation}
 where $\{\hat{M}_i\}$ is an orthonormal basis of the model subspace.
In terms of these new superoperators, the most general
 truncation that conserves $\|\hat{X}_M(\lambda)\|$ is 
\begin{subequations} \begin{align}
 \widehat{\mathcal{T}}(\lambda) &= k \widehat{\mathcal{P}}_M + \widehat{\mathcal{P}}_M \widehat{\mathcal{C}}[\hat{A}(\lambda)] \widehat{\mathcal{K}}(\lambda) \\
 \widehat{\mathcal{A}}(\lambda) &= \widehat{\mathcal{T}}(\lambda) \widehat{\mathcal{C}}[\hat{A}(\lambda)] \widehat{\mathcal{P}}_M,
\end{align} \end{subequations}
 where $k$ is real and $\widehat{\mathcal{K}}(\lambda)$ is anti-Hermitian.

Simple error minimization arguments complete the specification of $\widehat{\mathcal{T}}(\lambda)$.
The residual error of the truncated transformation is
\begin{align} \label{Rflow}
 \frac{d}{d\lambda} \hat{X}_R(\lambda) = & [\widehat{\mathcal{I}} - \widehat{\mathcal{T}}(\lambda)] [ \hat{A}(\lambda) , \hat{X}_M(\lambda) ]+ [\hat{A}(\lambda), \hat{X}_R(\lambda)],
\end{align}
 with the identity superoperator $\widehat{\mathcal{I}}$ and
 initial condition $\hat{X}_R(0) = \hat{X} - \hat{X}_M(0)$.
The second term just rotates the residual and is cancelled by considered a prerotated form,
 $\hat{X}_{R2}(\lambda) = \hat{U}(\lambda) \hat{X}_R(\lambda) \hat{U}^\dag(\lambda)$, defined by
\begin{equation} \label{Uflow}
 \frac{d}{d\lambda} \hat{U}(\lambda) = - \hat{U}(\lambda) \hat{A}(\lambda) \ \ \mathrm{and} \ \ \hat{U}(0) = \hat{I}.
\end{equation}
Assuming the details of $\hat{X}_R(\lambda)$ are unknown, the best strategy
 for minimizing $\|\hat{X}_R(\lambda)\|$ is to minimize
\begin{align} \label{Rgrow}
 \| d \hat{X}_{R2}(\lambda) / d\lambda \|^2 = & \|(\widehat{\mathcal{I}} - \widehat{\mathcal{P}}_M) [ \hat{A}(\lambda) , \hat{X}_M(\lambda) ]\|^2 \\
 & + \| [\widehat{\mathcal{P}}_M - \widehat{\mathcal{T}}(\lambda)] [ \hat{A}(\lambda) , \hat{X}_M(\lambda) ]] \|^2. \notag
\end{align}
The growth of error is minimized by a unique choice of truncation,
 $\widehat{\mathcal{T}}(\lambda) = \widehat{\mathcal{P}}_M$ ($k=1$ and $\widehat{\mathcal{K}}(\lambda) = 0$),
 which cancels the second term.
Operator truncation is now determined solely by the choice of model subspace.

By combining Eqs. (\ref{trunc_flow}), (\ref{Rflow}), and (\ref{Uflow}),
 the exact transformation of $\hat{X}$ can be partitioned into
\begin{equation} \label{error_form}
 \hat{U}^\dag(\lambda) \hat{X}\hat{U}(\lambda) = \hat{X}_M(\lambda) + \hat{X}_R(\lambda).
\end{equation}
To conform to the notation in Eq. (\ref{approx_diag}), the truncated
 eigenvalue decomposition of a Hermitian operator $\hat{H}$ is defined as
\begin{equation} \label{TED}
 \hat{U}^\dag(\infty) \hat{H} \hat{U}(\infty) = \hat{D} + \hat{R},
\end{equation}
 for a diagonal operator $\hat{D} = \hat{H}_M(\infty)$
 and a residual error $\hat{R} = \hat{H}_R(\infty)$.
The rest of the section is devoted to the theoretical issues of
 existence and uniqueness of solutions and the technical issues
 of efficient computability and error minimization.

The diagonalization of a Hermitian operator places special emphasis
 on Hermitian and diagonal operators.
Truncation should preserve both of these properties.
This can be enforced with restrictions on the model subspace.
If $\hat{X}$ is in the subspace, then $\hat{X}^\dag$ must also be in the subspace.
The model subspace must also be completely separable into
 a purely diagonal and purely off-diagonal subspace.
Projection superoperators can separate diagonal from off-diagonal (``coupling'') operators,
\begin{subequations} \begin{align}
 \widehat{\mathcal{P}}^D &= \sum_i |i\rangle \langle i | \otimes |i\rangle \langle i| \\
 \widehat{\mathcal{P}}^C &= \sum_{i \neq j} |i\rangle \langle j | \otimes |i\rangle \langle j|,
\end{align} \end{subequations}
 where $\{|i\rangle\}$ is an orthonormal basis of states that defines diagonality.
The model projector can be separated into
 $\widehat{\mathcal{P}}_M = \widehat{\mathcal{P}}_M^D + \widehat{\mathcal{P}}_M^C$,
 where $\widehat{\mathcal{P}}_M^D = \widehat{\mathcal{P}}_M \widehat{\mathcal{P}}^D$ 
 and $\widehat{\mathcal{P}}_M^C = \widehat{\mathcal{P}}_M \widehat{\mathcal{P}}^C$.
Model operators can similarly be split into
 $\hat{X}_M^D(\lambda) = \widehat{\mathcal{P}}^D \hat{X}_M(\lambda)$ and
 $\hat{X}_M^C(\lambda) = \widehat{\mathcal{P}}^C \hat{X}_M(\lambda)$.

\subsection{Existence and computability \label{existence}}

The most straightforward way to prove the existence
 of a solution to Eq. (\ref{TED}) is to construct one.
This simultaneously demonstrates that solutions are computable
 and lays the groundwork for a solution method.
Construction of a solution is guided by a convergence metric,
\begin{equation}
 \Omega(\lambda) = \frac{1}{2}\|\hat{H}_M^C(\lambda)\|^2,
\end{equation}
 which approaches zero as Eq. (\ref{TED}) is satisfied.
A solution exists if it is possible to choose $\hat{A}(\lambda)$
 to monotonically reduce $\Omega(\lambda)$ to zero.
This places important constraints on the generator subspace
 that contains $\hat{A}(\lambda)$.

It is almost always possible to reduce $\Omega(\lambda)$
 by enforcing $d\Omega(\lambda)/d\lambda < 0$.
The first derivative of $\Omega(\lambda)$ is
\begin{equation}
 \frac{d}{d\lambda} \Omega(\lambda) = - \langle [\hat{H}_M^D (\lambda), \hat{H}_M^C(\lambda)],\hat{A}(\lambda) \rangle.
\end{equation}
To maximize the rate of decrease of $\Omega(\lambda)$, the two operators
 in the inner product should be made as close to parallel as possible.
This criterion is limited by the restriction of $\hat{A}(\lambda)$
 to the generator subspace, which is enforced with a projection
 superoperator defined by an orthonormal basis $\{\hat{G}_i\}$
 of the generator subspace,
 \begin{equation}
 \widehat{\mathcal{P}}_G = \sum_i \hat{G}_i \otimes \hat{G}_i.
\end{equation}
The most parallel $\hat{A}(\lambda)$ is
\begin{equation} \label{steepA}
 \hat{A}(\lambda) = \widehat{\mathcal{P}}_G [\hat{H}_M^D (\lambda), \hat{H}_M^C(\lambda)].
\end{equation}
Without truncation, this procedure produces anti-Hermitian, off-diagonal operators.
These properties are preserved by truncation if the generator subspace
 is similarly restricted to contain only anti-Hermitian, off-diagonal operators.

In some cases, diagonalization can get stuck at a non-diagonal fixed point
 with $d\Omega(\lambda)/d\lambda = 0$ and $\Omega(\lambda) \neq 0$.
This occurs when $\hat{A}(\lambda) = 0$ and $\hat{H}_M^C(\lambda) \neq 0$ in Eq. (\ref{steepA})
 and signifies that $\hat{H}_M^C(\lambda)$ is contained in the right null space of
 $\widehat{\mathcal{P}}_G \widehat{\mathcal{C}}[\hat{H}_M^D (\lambda)] \widehat{\mathcal{P}}_M^C$.
To further reduce $\Omega(\lambda)$, the second derivative must be considered,
\begin{align}
 \frac{d^2}{d\lambda^2} \Omega(\lambda) = & - \left\langle [\hat{H}_M^D (\lambda), \hat{H}_M^C(\lambda)], \frac{d}{d\lambda}\hat{A}(\lambda) \right\rangle \notag \\
 & + \langle \widehat{\mathcal{P}}_M [\hat{H}_M^D(\lambda),\hat{A}(\lambda)] , [\hat{H}_M^C(\lambda),\hat{A}(\lambda)]\rangle \notag \\
 & + \| \widehat{\mathcal{P}}_M [\hat{H}_M^D(\lambda),\hat{A}(\lambda)] \|^2 \notag \\
 & - \| \widehat{\mathcal{P}}_M^D [\hat{H}_M^C(\lambda),\hat{A}(\lambda)] \|^2.
\end{align}
Monotonicity requires $d^2\Omega(\lambda) / d \lambda^2 \le 0$, but
 further reduction of $\Omega(\lambda)$ requires
 either $d^2\Omega(\lambda) / d \lambda^2 < 0$ or the consideration
 of even higher derivatives of $\Omega(\lambda)$.

To guarantee that $\Omega(\lambda)$ can be reduced to zero, the effects of the null space of
 $\widehat{\mathcal{P}}_G \widehat{\mathcal{C}}[\hat{H}_M^D (\lambda)] \widehat{\mathcal{P}}_M^C$
 have to be addressed.
If this superoperator's domain is restricted to the Hermitian off-diagonal model subspace
 and the range to the generator subspace, then constraining the dimensions of the
 subspaces to be equal is a necessary condition for the absence of a null space.
If a null space remains, then the right null space has a corresponding left null space.
When $d\Omega(\lambda)/d\lambda = 0$, the second derivative of $\Omega(\lambda)$
 can be made non-positive by restricting $\hat{A}(\lambda)$ to the left null space,
\begin{equation} \label{null_deriv2}
 \frac{d^2}{d\lambda^2} \Omega(\lambda) = - \| \widehat{\mathcal{P}}_M^D [\hat{H}_M^C(\lambda),\hat{A}(\lambda)] \|^2.
\end{equation}
The $d\Omega(\lambda)/d\lambda = 0$ fixed point is unstable
 if the second derivative can be made negative and nonzero.
A sufficient existence criteria is that for any $\hat{H}_M^C(\lambda)$ in the right null space of
 $\widehat{\mathcal{P}}_G \widehat{\mathcal{C}}[\hat{H}_M^D (\lambda)] \widehat{\mathcal{P}}_M^C$
 there must exist an $\hat{A}(\lambda)$ in the left null space such that Eq. (\ref{null_deriv2}) is nonzero.

To demonstrate that the existence criteria is necessary,
 the untruncated case is considered.
Without truncation, there is analytic eigenvalue decomposition of $\widehat{\mathcal{C}}[\hat{H}_M^D(\lambda)]$,
\begin{equation} \label{comm_ev}
 \widehat{\mathcal{C}}[\hat{H}_M^D(\lambda)] |i\rangle \langle j| = \left(\langle i | \hat{H}_M^D(\lambda) |i\rangle - \langle j | \hat{H}_M^D(\lambda) |j\rangle \right) |i\rangle \langle j|.
\end{equation}
The off-diagonal eigenoperators naturally come in pairs with eigenvalues of opposite sign,
 $\{|i\rangle \langle j|,|j\rangle \langle i|\}$.
When $\langle i | \hat{H}_M^D(\lambda) |i\rangle = \langle j | \hat{H}_M^D(\lambda) |j\rangle$,
 the eigenoperator pairs are degenerate and
 in the null space of $\widehat{\mathcal{C}}[\hat{H}_M^D(\lambda)]$.
This signifies a case where $d\Omega(\lambda)/d\lambda = 0$ and $\Omega(\lambda) \neq 0$
 if $\hat{H}_M^C(\lambda) \propto |i\rangle \langle j| + |j\rangle \langle i|$.
To satisfy the existence criteria for this case,
 the choice $\hat{A}(\lambda) \propto |i\rangle \langle j|-|j\rangle \langle i|$
 makes Eq. (\ref{null_deriv2}) nonzero.

A unique solution is defined by Eq. (\ref{steepA}) if the derivative
 of $\Omega(\lambda)$ remains nonzero until the problem is solved.
Uniqueness is lost when $d\Omega(\lambda)/d\lambda = 0$ because any
 $\hat{A}(\lambda)$ that satisfies the existence criteria has an arbitrary sign.
Each choice of sign leads to a different solution.
This almost always unique solution minimizes $\hat{R}$ in Eq. (\ref{TED})
 in a weak and indirect way.
The growth of truncation errors in Eq. (\ref{Rgrow}) is proportional
 to $\|\hat{A}(\lambda)\|$, and a smaller total error results from less transformation.
Orthogonal components could be added to Eq. (\ref{steepA}) to produce different solutions,
 but this increases $\|\hat{A}(\lambda)\|$ and thus truncation errors
 without affecting convergence to a solution as measured by $d\Omega(\lambda)/d\lambda$.
While a more sophisticated theory may be possible, this simple choice of solution
 establishes both uniqueness and error minimization for solutions
 to Eq. (\ref{TED}) in a weak but practical form.

\subsection{Efficient solution method \label{eigen_method}}

The continuous minimization of $\Omega(\lambda)$ can be used to construct
 solutions to Eq. (\ref{TED}), but it is not the most efficient method.
Numerical evolution of a differential equation is required to calculate $\hat{H}_M(\lambda)$.
$\hat{A}(\lambda)$ has to be stored at many $\lambda$ values 
 to transform operators after $\hat{H}$ has been diagonalized.
These problems are avoided if $\hat{A}(\lambda)$ is restricted to a piecewise
 constant function defined by a finite set of generators $\{\hat{A}_i\}$.
The continuous transformation in Eq. (\ref{trunc_flow}) can be analytically integrated
 over each constant generator
 to produce a sequence of unitary superoperator exponentials,
\begin{equation} \label{integrated_flow}
 \hat{X}_{M,i} = \exp ( \widehat{\mathcal{A}}_i ) \hat{X}_{M,i-1},
\end{equation}
 with $\hat{X}_{M,0} = \widehat{\mathcal{P}}_M \hat{X}$
 and $\widehat{\mathcal{A}}_i = \widehat{\mathcal{P}}_M \widehat{\mathcal{C}}[\hat{A}_i] \widehat{\mathcal{P}}_M$.
Some many-fermion methods \cite{Taube.UCC,Yanai.canonical.transformation,Bartlett.CCreview}
 use a single exponential for transformations, but section \ref{existence}
 cannot guarantee solutions in this case.
Efficiency is improved by reducing the cost of calculating each $\hat{A}_i$
 and minimizing the number required to diagonalize $\hat{H}$.

Eq. (\ref{integrated_flow}) can be efficiently evaluated as a Taylor series cast in a recursive form,
\begin{subequations} \begin{align}
 \hat{Z}_j &= \frac{1}{j}\widehat{\mathcal{P}}_M [\hat{A}_i,\hat{Z}_{j-1}] \\
 \sum_{j=0}^{\infty} \hat{Z}_j &= \exp (\widehat{\mathcal{A}}_i) \hat{X}_{M,i-1},
\end{align} \end{subequations}
 with $\hat{Z}_0 = \hat{X}_{M,i-1}$.
If $\hat{X}_{M,i-1}$ is overwritten with the sum over $\hat{Z}_j$,
 then the total memory requirement of this process is the storage of
 $\hat{X}_{M,i-1}$, $\hat{A}_i$, $\hat{Z}_j$, and $\hat{Z}_{j-1}$.
The calculation is terminated with a desired accuracy is reached,
 as estimated by \textit{a posteriori} and \textit{a priori} error bounds,
\begin{subequations} \label{exp_bounds} \begin{align}
 \left\| \sum_{j=d+1}^{\infty}\hat{Z}_j \right\| &\le d! \frac{a_{d}(\|\widehat{\mathcal{A}}_i\|)}{(\|\widehat{\mathcal{A}}_i\|)^{d}}  \| \hat{Z}_{d} \|\\
 &\le a_{d}(\|\widehat{\mathcal{A}}_i\|) \| \hat{X}_{M,i-1} \|.
\end{align} \end{subequations}
$a_d(x)$ bounds the error of a finite Taylor series approximation of the imaginary exponential,
\begin{equation}
  a_d(x) = \left| \exp(i x) - \sum_{j=0}^{d} \frac{(i x)^j}{j!} \right| .
\end{equation}
Eq. (\ref{exp_bounds}) is derived using the superoperator spectral norm
 and a superoperator function inequality,
\begin{subequations} \begin{align} \label{func_bound}
 \| f(\widehat{\mathcal{A}}_j) \hat{X} \| & \le \|\hat{X}\| \max_{|x| \le \| \widehat{\mathcal{A}}_j \|} |f(i x)|, \\
 \|\widehat{\mathcal{A}}_i\| & = \max_{\hat{Y}} \frac{\| \widehat{\mathcal{A}}_i \hat{Y} \|}{\|\hat{Y}\|}.
\end{align} \end{subequations}
$\|\widehat{\mathcal{A}}_i\|$ can be efficiently calculated by restricting
 $\hat{Y}$ to the model subspace and using the Lanczos method \cite{Parlett.max_ev}.

To guarantee the existence of solutions as in section \ref{existence},
 a method must calculate the $\hat{A}_i$ sequentially by reducing
 a discrete analogue of $\Omega(\lambda)$,
\begin{equation} \label{discrete_functional}
 \Omega_i[\hat{A}_i] = \frac{1}{2}\| \widehat{\mathcal{P}}^C \exp(\widehat{\mathcal{A}}_i) \hat{H}_{M,i-1} \|^2.
\end{equation}
Direct minimization of $\Omega_i[\hat{A}_i]$ requires the calculation
 of an accurate gradient, which is prohibitively expensive for large $\|\widehat{\mathcal{A}}_i\|$.
The gradient of $\Omega_i[\hat{A}_i]$ can be avoided by constructing
 a cheaper bounding functional, $\Omega_i[\hat{A}_i] \le \Lambda_i[\hat{A}_i]$.
Minimization of $\Lambda_i[\hat{A}_i]$ approximately minimizes $\Omega_i[\hat{A}_i]$.
With this strategy, the exponential of $\widehat{\mathcal{A}}_i$
 only needs to be accurately evaluated once per $i$ to calculate $\hat{H}_{M,i}$.
The tightness of the $\Lambda_i[\hat{A}_i]$ bound
 will determine the number of generators required to solve Eq. (\ref{TED}).
Specifically, if $\Lambda_i[\hat{A}_i] - \Omega_i[\hat{A}_i] \propto \|\widehat{\mathcal{A}}_i\|^d$ for small $\|\widehat{\mathcal{A}}_i\|$,
 then the asymptotic convergence will be $\Omega_i[\hat{A}_i] \propto \Omega_{i-1}^d[\hat{A}_{i-1}]$.
The only limitation on the choice of $\Lambda_i[\hat{A}_i]$ is that it must match
 $\Omega_i[\hat{A}_i]$ up to second order in $\hat{A}_i$ to correctly treat
 the cases in section \ref{existence} where $d^2\Omega(\lambda)/d\lambda^2$
 is required to calculate $\hat{A}(\lambda)$.

Ultimately, the choice of $\Lambda_i[\hat{A}_i]$ must be guided by the
 specific details of a particular application and computing environment.
Some examples of bounds on $\Omega_i[\hat{A}_i]$ are
\begin{equation} \label{many_bounds}
 \sqrt{2 \Omega_i[\hat{A}_i]} \le \left\| \widehat{\mathcal{P}}^C \sum_{j=0}^d \frac{\widehat{\mathcal{A}}_i^j}{j!} \hat{H}_{M,i-1} \right\| + a_d(\|\widehat{\mathcal{A}}_i\|) \|\hat{H}_{M,i-1}\|,
\end{equation}
 derived by splitting the exponential Taylor series
 with the triangle inequality and applying Eq. (\ref{func_bound}).
If one of these bounds is minimized,
 $a_d(\|\widehat{\mathcal{A}}_i\|)$ effectively acts as a
 Lagrange multiplier to constrain the size of $\|\widehat{\mathcal{A}}_i\|$
 and thus reduce unnecessary transformations and their associated truncation errors.
Whatever the choice of $\Lambda_i[\hat{A}_i]$, standard methods can be used for functional minimization.
Convergence of the minimization procedure will depend on
 the condition number of the Hessian matrix and
 the ability to effectively precondition the gradient.

\section{Application to fermions \label{for_fermions}}

Several ingredients are required to apply the theory developed
 in section \ref{eigen_theory} to many-fermion systems.
First, new notation is introduced to simplify many-fermion operator algebra.
Second, model and generator subspaces are chosen
 to satisfy the existence criteria in section \ref{existence}.
Third, a bounding functional $\Lambda_i[\hat{A}_i]$ and
 its gradient preconditioner are specified
 to satisfy the requirements in section \ref{eigen_method}.

A few basic but non-standard occupation vector operations are used throughout this section.
Vector-valued operations are modular addition and three set-theoretic operations,
 defined by their components as
\begin{subequations} \begin{align}
 [\mathbf{f}\oplus\mathbf{g}]_i &= (f_i + g_i) \bmod 2 \\
 [\mathbf{f}\cap\mathbf{g}]_i &= f_i g_i \\ 
 [\mathbf{f}\setminus\mathbf{g}]_i &= f_i - f_i g_i \\
 [\mathbf{f}\cup\mathbf{g}]_i &= f_i + g_i - f_i g_i.
\end{align} \end{subequations}
An occupation vector norm is defined as $\|\mathbf{x}\| = \sum_i |x_i|$.
Also useful is a fermion sign function,
\begin{equation}
 s(\mathbf{f},\mathbf{g}) = (-1)^{\sum_i (f_i \sum_{j<i} g_j)},
\end{equation}
 which obeys several useful identities,
\begin{subequations} \begin{align}
 s(\mathbf{f},\mathbf{g}) s(\mathbf{h},\mathbf{g}) &= s(\mathbf{f} \oplus \mathbf{h},\mathbf{g}) \\
 s(\mathbf{f},\mathbf{g}) s(\mathbf{f},\mathbf{h}) &= s(\mathbf{f},\mathbf{g} \oplus \mathbf{h}) \\
 s(\mathbf{f},\mathbf{f}) &= (-1)^{\lfloor \|\mathbf{f}\| / 2\rfloor} \\
 s(\mathbf{f},\mathbf{g}) s(\mathbf{g},\mathbf{f}) &= (-1)^{\|\mathbf{f}\| \|\mathbf{g}\| - \mathbf{f} \cdot \mathbf{g}},
\end{align} \end{subequations}
 and accounts for all sign changes resulting from fermion anti-commutations.

Operator algebra is simplified by indexing operator basis elements
 with pairs of occupation vectors.
A simple example of such an operator is an
 outer product of Slater determinants, $|\mathbf{f}\rangle\langle\mathbf{g}|$.
However, physical many-fermion operators such as in Eq. (\ref{hamiltonian})
 are compactly represented with products of elementary fermion operators
 and not Slater determinant outer products.
Two types of operator basis elements of this form are considered,
\begin{subequations} \begin{align} \label{B_def}
  \hat{B}(\mathbf{f},\mathbf{g}) = & i^{\lfloor \|\mathbf{f}\oplus\mathbf{g}\| / 2\rfloor \bmod 2} \notag \\
  &\times \prod_j [(1-f_j)(1-g_j) + (1 - 2 \hat{n}_j) f_j g_j \notag \\ 
  & \qquad \ \ + (\hat{c}_j^\dag + \hat{c}_j) f_j (1-g_j) \notag \\
  & \qquad \ \ + i (\hat{c}_j^\dag - \hat{c}_j) (1-f_j) g_j], \\
 \hat{C}(\mathbf{f},\mathbf{g})  = & \prod_j [ (1-f_j)(1-g_j) + (1 - 2 \hat{n}_j) f_j g_j  \notag \\
 & \quad \ \ + \hat{c}_j^\dag f_j (1-g_j) + \hat{c}_j (1-f_j) g_j],
\end{align} \end{subequations}
 each with distinct advantages and disadvantages.

$\hat{B}(\mathbf{f},\mathbf{g})$ and $\hat{C}(\mathbf{f},\mathbf{g})$ share several basic properties.
They are both off-diagonal for $\mathbf{f}\neq\mathbf{g}$
 and diagonal for $\mathbf{f}=\mathbf{g}$.
Also, $\hat{B}(\mathbf{f},\mathbf{f}) = \hat{C}(\mathbf{f},\mathbf{f})$.
Each set of elements is trace-orthogonal and both have simple normalizations,
\begin{equation}
 \| \hat{B}(\mathbf{f},\mathbf{g}) \|^2 = 2^n \ \ \mathrm{and} \ \ \| \hat{C}(\mathbf{f},\mathbf{g}) \|^2 = 2^{n-\|\mathbf{f}\oplus\mathbf{g}\|},
\end{equation}
 where $n$ is the total number of fermion degrees of freedom.
Operators can be decomposed in either basis,
\begin{subequations} \label{BC_trans} \begin{align} 
 \hat{X} &= 2^{-n} \sum_{\mathbf{f},\mathbf{g}} \mathrm{Tr}[\hat{B}(\mathbf{f},\mathbf{g})\hat{X}] \hat{B}(\mathbf{f},\mathbf{g}) \\
 \hat{X} &= 2^{-n} \sum_{\mathbf{f},\mathbf{g}} 2^{\|\mathbf{f}\oplus\mathbf{g}\|} \mathrm{Tr}[\hat{C}^\dag(\mathbf{f},\mathbf{g}) \hat{X}] \hat{C}(\mathbf{f},\mathbf{g}).
\end{align} \end{subequations}
Basis transformations can be calculated by representing the elements of one basis in another basis,
\begin{subequations} \label{B_to_C} \begin{align}
 \hat{C}(\mathbf{f},\mathbf{g}) = & i^{-\lfloor \|\mathbf{f}\oplus\mathbf{g}\| / 2\rfloor \bmod 2 + \|\mathbf{g}\setminus\mathbf{f}\|} 2^{-\|\mathbf{f}\oplus\mathbf{g}\|} \notag \\
 & \times \sum_{\mathbf{h} \setminus (\mathbf{f}\oplus\mathbf{g}) = 0} i^{-(\mathbf{f}\oplus\mathbf{g})\cdot\mathbf{h}} \hat{B}(\mathbf{f}\oplus\mathbf{h},\mathbf{g}\oplus\mathbf{h}) \\
 \hat{B}(\mathbf{f},\mathbf{g}) = & i^{\lfloor \|\mathbf{f}\oplus\mathbf{g}\| / 2\rfloor \bmod 2 - \|\mathbf{g}\setminus\mathbf{f}\|} \notag \\
 & \times \sum_{\mathbf{h} \setminus (\mathbf{f}\oplus\mathbf{g}) = 0} (-1)^{\mathbf{g}\cdot\mathbf{h}} \hat{C}(\mathbf{f}\oplus\mathbf{h},\mathbf{g}\oplus\mathbf{h}). 
\end{align} \end{subequations}
Further properties of the operators deviate.

The $\hat{B}(\mathbf{f},\mathbf{g})$ operators are Hermitian and unitary
 and have simple algebraic properties.
The action of $\hat{B}(\mathbf{f},\mathbf{g})$ on a Slater determinant produces 
 another Slater determinant with an $i^m$ phase factor,
\begin{subequations} \begin{align}
 \hat{B}(\mathbf{f},\mathbf{g}) |\mathbf{h}\rangle = &\theta(\mathbf{f},\mathbf{g},\mathbf{h}) |\mathbf{f}\oplus\mathbf{g}\oplus\mathbf{h}\rangle \\
 \theta(\mathbf{f},\mathbf{g},\mathbf{h}) = &(-1)^{\mathbf{g}\cdot\mathbf{h}} s(\mathbf{f}\oplus\mathbf{g},\mathbf{h}) \notag \\
 &\times i^{ \lfloor \|\mathbf{f}\oplus\mathbf{g}\| / 2\rfloor \bmod 2 + \|\mathbf{g}\setminus\mathbf{f}\|}.
\end{align} \end{subequations}
The product of two $\hat{B}(\mathbf{f},\mathbf{g})$ basis operators
 is a single other basis operator with another $i^m$ phase factor,
\begin{subequations} \begin{align}
 \hat{B}(\mathbf{f},\mathbf{g}) \hat{B}(\mathbf{h},\mathbf{k}) = & \phi(\mathbf{f},\mathbf{g},\mathbf{h},\mathbf{k}) \hat{B}(\mathbf{f}\oplus\mathbf{h},\mathbf{g}\oplus\mathbf{k}) \\
 \phi(\mathbf{f},\mathbf{g},\mathbf{h},\mathbf{k}) = & s(\mathbf{f}\oplus\mathbf{g},\mathbf{h}\oplus\mathbf{k}) (-1)^{(\mathbf{g}\setminus\mathbf{f})\cdot(\mathbf{h}\setminus\mathbf{k})} \notag \\
 &\times (-1)^{(\mathbf{f}\cap\mathbf{g})\cdot(\mathbf{k}\setminus\mathbf{h})+(\mathbf{f}\setminus\mathbf{g})\cdot(\mathbf{h}\cap\mathbf{k})} \notag \\
 &\times i^{[(\mathbf{f}\oplus\mathbf{g})\cdot(\mathbf{h}\oplus\mathbf{k}) - \|\mathbf{f}\oplus\mathbf{g}\| \|\mathbf{h}\oplus\mathbf{k}\|] \bmod 2} \notag \\
 &\times i^{\|(\mathbf{f}\cap\mathbf{k})\oplus(\mathbf{g}\cap\mathbf{h})\|}.
\end{align} \end{subequations}
The commutator formula is equally simple,
\begin{align}
 [\hat{B}(\mathbf{f},\mathbf{g}) , \hat{B}(\mathbf{h},\mathbf{k})] = & 2 i \textrm{Im}\phi(\mathbf{f},\mathbf{g},\mathbf{h},\mathbf{k}) \hat{B}(\mathbf{f}\oplus\mathbf{h},\mathbf{g}\oplus\mathbf{k}).
\end{align}
The algebra of the $\hat{B}(\mathbf{f},\mathbf{g})$ operators bears similarities to the Pauli matrices
 and they might be considered as a many-fermion analogue.

The important advantage of $\hat{C}(\mathbf{f},\mathbf{g})$ over $\hat{B}(\mathbf{f},\mathbf{g})$
 is its ability to exploit 1-fermion symmetries.
These symmetries arise from 1-fermion invariant operators that commute with a Hamiltonian, $[\hat{S},\hat{H}] = 0$.
Without loss of generality, $\hat{S}$ is Hermitian.
The vanishing commutator allows $\hat{S}$ and $\hat{H}$ to be simultaneously diagonalized.
The fermion operators can be chosen to diagonalize $\hat{S}$,
 resulting in the simple commutation relations
\begin{equation}
 [\hat{S},\hat{c}_i] = s_i \hat{c}_i.
\end{equation}
Examples of common 1-fermion invariant operators are total fermion number,
 total fermion spin, lattice vector translations, and point group operations.
The $\hat{C}(\mathbf{f},\mathbf{g})$ operators retain the simple commutation relations,
\begin{equation}
 [\hat{S},\hat{C}(\mathbf{f},\mathbf{g})] = \hat{C}(\mathbf{f},\mathbf{g}) \sum_i s_i (g_i - f_i) .
\end{equation}
Only $\hat{C}(\mathbf{f},\mathbf{g})$ operators that commute with $\hat{S}$ are required
 to represent $\hat{H}$ and the symmetry-preserving generators $\hat{A}$ that diagonalize it.

The disadvantage of the $\hat{C}(\mathbf{f},\mathbf{g})$ operators
 is their more complicated operator algebra.
They are not Hermitian and are related to their Hermitian conjugates by
\begin{equation}
 \hat{C}^\dag(\mathbf{f},\mathbf{g}) = (-1)^{\lfloor\|\mathbf{f}\oplus\mathbf{g}\|/2\rfloor}\hat{C}(\mathbf{g},\mathbf{f}).
\end{equation}
The action of $\hat{C}(\mathbf{f},\mathbf{g})$ on Fock states can now produce zero,
\begin{align}
 \hat{C}(\mathbf{f},\mathbf{g}) |\mathbf{h}\rangle = & v(\mathbf{h}-\mathbf{g}+\mathbf{f}) (-1)^{(\mathbf{f}\cap\mathbf{g})\cdot\mathbf{h}} \notag \\
 &\times s(\mathbf{f}\oplus\mathbf{g},\mathbf{h}) |\mathbf{f}\oplus\mathbf{g}\oplus\mathbf{h}\rangle,
\end{align}
 encoded in an occupation vector validity function,
\begin{equation}
 v(\mathbf{x}) = \left \{ \begin{array}{ll} 1, & \mathbf{x} \in \{0,1\}^n \\ 0, & \mathrm{otherwise} \end{array} \right . .
\end{equation}
The product of two basis operators is no longer always a single basis operator,
\begin{widetext} \begin{align}
 \hat{C}(\mathbf{f},\mathbf{g}) \hat{C}(\mathbf{h},\mathbf{k}) = & v(\mathbf{f} - \mathbf{g} + \mathbf{h} - \mathbf{k}) s(\mathbf{f}\oplus\mathbf{g},\mathbf{h}\oplus\mathbf{k}) (-1)^{(\mathbf{f}\oplus\mathbf{k})\cdot(\mathbf{g}\cap\mathbf{h})}2^{-(\mathbf{f}\oplus\mathbf{g})\cdot(\mathbf{h}\oplus\mathbf{k})} \notag \\
 & \times \sum_{\mathbf{z} \setminus [(\mathbf{f}\oplus\mathbf{g})\cap(\mathbf{h}\oplus\mathbf{k})] = 0}  (-1)^{\|\mathbf{z}\setminus(\mathbf{g}\cap\mathbf{h})\|} \hat{C}(\mathbf{x}\oplus\mathbf{z},\mathbf{y}\oplus\mathbf{z}) \\
 \mathrm{with} \ \ \mathbf{x} = & [(\mathbf{f}\setminus\mathbf{g})\setminus(\mathbf{k}\setminus\mathbf{h})] \oplus [(\mathbf{h}\setminus\mathbf{k})\setminus(\mathbf{g}\setminus\mathbf{f})] \oplus [(\mathbf{f}\cap\mathbf{g})\setminus(\mathbf{h}\cup\mathbf{k})] \oplus [(\mathbf{h}\cap\mathbf{k})\setminus(\mathbf{f}\cup\mathbf{g})] \notag \\
 \mathrm{and} \ \ \mathbf{y} = & [(\mathbf{g}\setminus\mathbf{f})\setminus(\mathbf{h}\setminus\mathbf{k})] \oplus [(\mathbf{k}\setminus\mathbf{h})\setminus(\mathbf{f}\setminus\mathbf{g})] \oplus [(\mathbf{f}\cap\mathbf{g})\setminus(\mathbf{h}\cup\mathbf{k})] \oplus [(\mathbf{h}\cap\mathbf{k})\setminus(\mathbf{f}\cup\mathbf{g})]. \notag
\end{align} \end{widetext}
In this arrangement, $\hat{C}(\mathbf{x},\mathbf{y})$ is the contribution
 with the smallest value of $\|(\mathbf{x}\oplus\mathbf{z})\cup(\mathbf{y}\oplus\mathbf{z})\|$.
The commutator formula is just two applications of the product formula
 and does not further simplify except to cancel some terms in the sum over $\mathbf{z}$
 and vanish when
\begin{align}
  (\mathbf{f}\oplus\mathbf{g})\cdot(\mathbf{h}\oplus\mathbf{k}) & = 0 \ \ \mathrm{and} \\
 [ (\mathbf{f}\oplus\mathbf{k})\cdot(\mathbf{g}\cap\mathbf{h})&+(\mathbf{g}\oplus\mathbf{h})\cdot(\mathbf{f}\cap\mathbf{k}) \notag \\
 & + \|\mathbf{f}\oplus\mathbf{g}\| \|\mathbf{h}\oplus\mathbf{k}\|] \bmod 2 = 0. \notag
\end{align}

\subsection{Model and generator subspaces}

The model subspace of a many-fermion system should contain
 all operators necessary for an accurate physical description
 of the Hamiltonian as it is transformed to a diagonal form.
It is generally observed that basis operators containing fewer elementary
 fermion operators have more physical importance.
In systems where the physics is geometrically local, geometric constraints
 may also determine the importance of basis operators.
Careful study might reveal further crucial system-specific sets of basis operators,
 distinct from either general criterion.
To allow for all these possibilities, the model subspace
 is defined by an allowed set of occupation vectors $V$ as
\begin{align} \label{V.sparse}
 \{ \hat{B}(\mathbf{f},\mathbf{g}) : \mathbf{f}\cup\mathbf{g} \in V \}, \ \ \ \mathbf{f} \in V \implies \mathbf{f}\setminus\mathbf{x} \in V.
\end{align}
$\hat{B}(\mathbf{f},\mathbf{g})$ and $\hat{C}(\mathbf{f},\mathbf{g})$
 are interchangeable in this definition.
The constraints on $V$ are minimal and it is straightforward
 to expand any set to satisfy them.
Operators in the span of this basis are defined to be \textit{$V$-sparse}.

The generator subspace is chosen to contain all off-diagonal
 and anti-Hermitian operators in the model subspace.
In terms of $\hat{B}(\mathbf{f},\mathbf{g})$, this means basis elements
 of the form $i \hat{B}(\mathbf{f},\mathbf{g})$, $\mathbf{f}\neq\mathbf{g}$.
In terms of $\hat{C}(\mathbf{f},\mathbf{g})$, this means basis elements of the form
 $\hat{C}(\mathbf{f},\mathbf{g}) - \hat{C}^\dag(\mathbf{f},\mathbf{g})$ for a real operator space.
For a complex operator space, the subspace must also include
 $i[\hat{C}(\mathbf{f},\mathbf{g}) + \hat{C}^\dag(\mathbf{f},\mathbf{g})]$, $\mathbf{f}\neq\mathbf{g}$.
This choice minimizes the truncation errors in a weak, ``greedy'' way.
For a given $\mathbf{f}$ in $V$, if a Hamiltonian contains only terms of the form
 $\hat{B}(\mathbf{g},\mathbf{h})$ with $(\mathbf{g}\cup\mathbf{h})\setminus\mathbf{f} = 0$,
 then diagonalization can be performed exactly.

The use of $\hat{B}(\mathbf{f},\mathbf{g})$ or $\hat{C}(\mathbf{f},\mathbf{g})$
 as both an operator basis and subspace basis simplifies operator truncation.
Since the basis elements are trace-orthogonal, projections using $\widehat{\mathcal{P}}_M$ or $\widehat{\mathcal{P}}_G$
 just discard elements not in the subspace.
This can be physically interpretted as a form of normal ordering based truncation.
The standard normal ordering rules for a reference Slater determinant $|\mathbf{z}\rangle$
 arrange all number operators into the form $\hat{n}_m - z_m$ as in
 $\hat{C}^{\mathbf{z}}(\mathbf{f},\mathbf{g})$ defined below in Eq. (\ref{normalC}).
If the reference is an ensemble of Slater determinants with statistically uncorrelated occupations,
 $z_m \in [0,1]$, the standard normal ordering rules still apply.
The $\hat{C}(\mathbf{f},\mathbf{g})$ operators and truncation derived in section \ref{eigen_theory}
 correspond to the infinite temperature thermal ensemble, $z_m = 1/2$,
 that equally weights all fermion configurations.
This prevents the physics from being biased by a choice of reference state.
For a truncation process meant to approximate the entire spectrum of a many-fermion system,
 it is unreasonable to expect any one reference state to be suitable for the description of all eigenstates.

The suggested choice of model and generator subspaces defined by $V$-sparsity
 satisfies all the criteria established in section \ref{existence}
 to guarantee Hamiltonian diagonalization.
The only criterion that requires discussion is the
 existence of non-zero values for Eq. (\ref{null_deriv2}).
The analysis is simplest in the $\hat{C}(\mathbf{f},\mathbf{g})$ basis
 because the truncated diagonal commutator
 $\widehat{\mathcal{P}}_G \widehat{\mathcal{C}}[\hat{H}_M^D] \widehat{\mathcal{P}}_M^C$
 preserves the vector $\mathbf{f}-\mathbf{g}$.
This property can be exploited by writing the Hamiltonian in the form
\begin{equation}
 \hat{H}_M = \hat{H}_M^D + \sum_{\mathbf{f}\cdot\mathbf{g} = 0} \hat{C}(\mathbf{f},\mathbf{g}) \hat{D}(\mathbf{f},\mathbf{g}),
\end{equation}
 where $\hat{D}(\mathbf{f},\mathbf{g})$ are non-zero diagonal Hermitian operators that commute with $\hat{C}(\mathbf{f},\mathbf{g})$.
If $\hat{H}_M$ is in the right null space of $\widehat{\mathcal{P}}_G \widehat{\mathcal{C}}[\hat{H}_M^D] \widehat{\mathcal{P}}_M^C$,
 then each non-zero term of the form $\hat{C}(\mathbf{f},\mathbf{g}) \hat{D}(\mathbf{f},\mathbf{g})$
 is also in the null space.
The generator $\hat{A}$ can be chosen as any anti-Hermitian combination of these null operators.
If the generator is chosen to be $[\hat{C}(\mathbf{f},\mathbf{g}) - \hat{C}^\dag(\mathbf{f},\mathbf{g})] \hat{D}(\mathbf{f},\mathbf{g})$
 then the existence condition reduces to
\begin{equation} \label{specific_existence}
 \|\widehat{\mathcal{P}}_M \hat{D}^2(\mathbf{f},\mathbf{g}) [\hat{C}(\mathbf{f},\mathbf{g}),\hat{C}(\mathbf{g},\mathbf{f})]\| \neq 0.
\end{equation}
The left hand side of Eq. (\ref{specific_existence}) can be bounded from below by replacing 
 $\hat{D}^2(\mathbf{f},\mathbf{g})$ by its trace.
For $\hat{D}(\mathbf{f},\mathbf{g}) \neq 0$, the trace of
 $\hat{D}^2(\mathbf{f},\mathbf{g})$ is non-zero and can be ignored.
The remaining commutator is unaffected by the truncation and can be explicitly calculated as
\begin{subequations} \begin{align}
 \| [\hat{C}(\mathbf{f},\mathbf{g}),\hat{C}(\mathbf{g},\mathbf{f})] \| = & 2^{-\|\mathbf{f}\oplus\mathbf{g}\|} \sqrt{2^{n+1} b(\|\mathbf{f}\oplus\mathbf{g}\|)} \\
 b(i) = \sum_{j=0}^{\lfloor(i-1)/2\rfloor} & \frac{i !}{(i-2j-1) ! (2j+1)!}.
\end{align} \end{subequations}
The commutator always has a non-zero norm, which establishes that
 Eq. (\ref{specific_existence}) is satisfied.

\subsection{Bounding functional and preconditioner}

The bounding functional is chosen to match the form of Eq. (\ref{many_bounds})
 for $d = 2$, which is the simplest allowed functional of that form.
With this choice, the asymptotic convergence is $\|\hat{H}_{M,i}^C\| \propto \|\hat{H}_{M,i-1}^C\|^3$.
The functional can be written suggestively as
\begin{subequations} \begin{align}
 \Lambda_i[\hat{A}_i] = & \frac{1}{2} [ \| \hat{H}_{M,i-1}^C + \widehat{\mathcal{P}}_M^C [\hat{A}_i,\hat{H}'_{M,i-1}] \| \notag \\
 & \quad + a_2(\alpha\|\hat{A}_i\|) \|\hat{H}_{M,i-1}\| ]^2 \\
 \hat{H}'_{M,i-1} = & \hat{H}_{M,i-1} + \frac{1}{2} \widehat{\mathcal{P}}_M [\hat{A}_i,\hat{H}_{M,i-1}] \\
 \alpha = & \| \widehat{\mathcal{A}}_i \| / \|\hat{A}_i\|.
\end{align} \end{subequations}
This form enables the calculation of $\| \widehat{\mathcal{A}}_i \|$ to be weakly coupled
 to the minimization of $\Lambda_i[\hat{A}_i]$ if $\alpha$ has a weak dependence on $\hat{A}_i$.

A preconditioner is constructed by approximating the inverse Hessian of $\Lambda_i[\hat{A}_i]$.
This is straightforward in the untruncated case.
When diagonalization is converged, the Hessian reduces to $\widehat{\mathcal{C}}[\hat{D}]^2$.
A preconditioner that is exact in this limit is
\begin{subequations} \begin{align} \label{exact_preconditioner}
 \widehat{\mathcal{F}} = & \sum_{\mathbf{f} \neq \mathbf{g}} \frac{|\mathbf{f}\rangle\langle\mathbf{g}| \otimes |\mathbf{f}\rangle\langle\mathbf{g}|}{\Delta(\mathbf{f},\mathbf{g})}, \\
 \Delta(\mathbf{f},\mathbf{g}) = & (\langle\mathbf{f}|\hat{H}_{M,i-1}|\mathbf{f}\rangle - \langle\mathbf{g}|\hat{H}_{M,i-1}|\mathbf{g}\rangle)^2 \notag \\ & + 4 | \langle\mathbf{f}|\hat{H}_{M,i-1}|\mathbf{g}\rangle |^2 + \beta.
\end{align} \end{subequations}
This form includes approximate off-diagonal corrections using the quadratic formula
 and an extra uniform shift $\beta$.
The shift acts as either a tuning parameter
 to adjust the size of the preconditioned gradient or to approximate the effects of
 $a_2(\alpha\|\hat{A}_i\|)$ on the Hessian when $\|\hat{A}_i\|$ gets large.
(WIP) 

There is no natural extension of Eq. (\ref{exact_preconditioner}) to the truncated case
 using the $\hat{B}(\mathbf{f},\mathbf{g})$ or $\hat{C}(\mathbf{f},\mathbf{g})$ basis elements.
However, this preconditioner can be approximated by defining a basis $\hat{E}_V^{\mathbf{z}}(\mathbf{f},\mathbf{g})$
 of the $V$-sparse subspace that mimics the Slater determinant outer products,
\begin{equation}
 \langle \mathbf{f} \oplus \mathbf{z} | \hat{E}_V^{\mathbf{z}}(\mathbf{h},\mathbf{k}) | \mathbf{g} \oplus \mathbf{z} \rangle = \delta[\|\mathbf{f}-\mathbf{h}\|] \delta[\|\mathbf{g}-\mathbf{k}\|]
\end{equation}
 for $\mathbf{f}\cup\mathbf{g} \in V$ and $\mathbf{h}\cup\mathbf{k} \in V$.
$\delta[i]$ is the Kronecker delta function ($\delta[0] = 1$,$\delta[i\neq0] = 0$).
These operators are defined with respect to a reference Slater determinant $|\mathbf{z}\rangle$.
A $V$-sparse operator $\hat{X}_V$ can be decomposed in this basis using its matrix elements,
\begin{equation} \label{quasi-sparsity}
 \hat{X}_V = \sum_{\mathbf{f}\cup\mathbf{g} \in V} \langle \mathbf{f} \oplus \mathbf{z} | \hat{X}_V | \mathbf{g} \oplus \mathbf{z} \rangle \hat{E}_V^{\mathbf{z}}(\mathbf{f},\mathbf{g}).
\end{equation}
The analogue of Eq. (\ref{exact_preconditioner}) using $\hat{E}_V^{\mathbf{z}}(\mathbf{f},\mathbf{g})$ is
\begin{equation}
 \widehat{\mathcal{F}} = \sum_{\substack{\mathbf{f}\cup\mathbf{g} \in V \\ \mathbf{f} \neq \mathbf{g}}} \frac{\hat{E}_V^{\mathbf{z}}(\mathbf{f},\mathbf{g}) \otimes |\mathbf{f}\oplus\mathbf{z}\rangle\langle\mathbf{g}\oplus\mathbf{z}|}{\Delta(\mathbf{f}\oplus\mathbf{z},\mathbf{g}\oplus\mathbf{z})}.
\end{equation}
The choice of $\mathbf{z}$ is arbitrary, but it should not have
 a strong effect on the quality of the preconditioner.

An efficient explicit construction of $\hat{E}_V^{\mathbf{z}}(\mathbf{f},\mathbf{g})$
 requires the definition of an intermediate operator basis,
\begin{align} \label{normalC}
 \hat{C}^{\mathbf{z}}(\mathbf{f},\mathbf{g}) = & s(\mathbf{f}\oplus\mathbf{g},\mathbf{g}\oplus\mathbf{z}) \notag \\
 &\times \prod_i \{ (1-f_i)(1-g_i) \\
 & \qquad \quad + \hat{n}_i f_i g_i + (1-2 \hat{n}_i) z_i f_i g_i   \notag \\
 & \qquad \quad + \hat{c}_i^\dag [f_i (1-g_i) + z_i (g_i - f_i)] \notag \\
 & \qquad \quad + \hat{c}_i [(1-f_i) g_i + z_i (f_i - g_i)] \} \notag.
\end{align}
This is a variant of $\hat{C}(\mathbf{f},\mathbf{g})$
 that is normal ordered with respect to $|\mathbf{z}\rangle$.
$\hat{E}_V^{\mathbf{z}}(\mathbf{f},\mathbf{g})$ can be constructed by writing
 $|\mathbf{f}\oplus\mathbf{z}\rangle \langle \mathbf{g}\oplus\mathbf{z}|$
 in terms of $\hat{C}^{\mathbf{z}}(\mathbf{f},\mathbf{g})$ and projecting into the $V$-sparse subspace,
\begin{subequations} \label{C_to_E} \begin{align}
 \hat{E}_V^{\mathbf{z}}(\mathbf{f},\mathbf{g}) &= \sum_{\substack{(\mathbf{f}\cup\mathbf{g})\oplus\mathbf{h} \in V \\ (\mathbf{f}\cup\mathbf{g})\cdot\mathbf{h} = 0}} (-1)^{\|\mathbf{h}\|}\hat{C}^{\mathbf{z}} (\mathbf{f}\oplus\mathbf{h},\mathbf{g}\oplus\mathbf{h}) \\
 \hat{C}^{\mathbf{z}}(\mathbf{f},\mathbf{g}) &= \sum_{\substack{(\mathbf{f}\cup\mathbf{g})\oplus\mathbf{h} \in V \\ (\mathbf{f}\cup\mathbf{g})\cdot\mathbf{h} = 0}} s(\mathbf{f}\oplus\mathbf{g},\mathbf{h})\hat{E}_V^{\mathbf{z}} (\mathbf{f}\oplus\mathbf{h},\mathbf{g}\oplus\mathbf{h}).
\end{align} \end{subequations}
The inverse transformation is calculated using Eq. (\ref{quasi-sparsity}).
The transformations between $\hat{E}_V^{\mathbf{z}}(\mathbf{f},\mathbf{g})$ and
 $\hat{C}^{\mathbf{z}}(\mathbf{f},\mathbf{g})$ are independent of $\mathbf{z}$.

A few remaining formulas are required to transform between
 the $\hat{B}(\mathbf{f},\mathbf{g})$ basis and
 $\hat{E}_V^{\mathbf{z}}(\mathbf{f},\mathbf{g})$ basis.
The missing intermediate steps are the transformations between
 $\hat{C}(\mathbf{f},\mathbf{g})$ and $\hat{C}^{\mathbf{z}}(\mathbf{f},\mathbf{g})$,
\begin{subequations} \label{C_to_Cz} \begin{align}
 \hat{C}(\mathbf{f},\mathbf{g}) = & (-1)^{(\mathbf{f}\cap\mathbf{g})\cdot\mathbf{z}} s(\mathbf{f}\oplus\mathbf{g},\mathbf{g}\oplus\mathbf{z}) \\
 & \times \sum_{\mathbf{h}\setminus(\mathbf{f}\cap\mathbf{g}) = 0} (-2)^{\|\mathbf{h}\|} s(\mathbf{f}\oplus\mathbf{g},\mathbf{h}) \notag \\
 & \qquad \qquad \quad \times \hat{C}^{\mathbf{z}} (\mathbf{f}'\oplus\mathbf{h},\mathbf{g}'\oplus\mathbf{h}) \notag \\
 \hat{C}^{\mathbf{z}}(\mathbf{f},\mathbf{g}) = & 2^{-\mathbf{f}\cdot\mathbf{g}} s(\mathbf{f}\oplus\mathbf{g},\mathbf{g}\oplus\mathbf{z}) \\
 & \times \sum_{\mathbf{h}\setminus(\mathbf{f}\cap\mathbf{g}) = 0} (-1)^{\|\mathbf{h}\setminus\mathbf{z}\|} \hat{C}(\mathbf{f}'\oplus\mathbf{h},\mathbf{g}'\oplus\mathbf{h}) \notag, \\
 \mathrm{with} \ \mathbf{f}' = & (\mathbf{f}\setminus\mathbf{z}) \oplus (\mathbf{g}\cap\mathbf{z}) \oplus (\mathbf{f}\cap\mathbf{g}) \\
 \mathrm{and} \ \mathbf{g}' = & (\mathbf{g}\setminus\mathbf{z}) \oplus (\mathbf{f}\cap\mathbf{z}) \oplus (\mathbf{f}\cap\mathbf{g}).
\end{align} \end{subequations}
The complete transformation is performed as a sequence of three intermediate steps:
 $\hat{B} \leftrightarrow \hat{C}$ using Eq. (\ref{B_to_C}),
 $\hat{C} \leftrightarrow \hat{C}^\mathbf{z}$ using Eq. (\ref{C_to_Cz}),
 and $\hat{C}^\mathbf{z} \leftrightarrow \hat{E}_V^\mathbf{z}$ using Eq. (\ref{C_to_E}).

\section{Truncated eigenfermion decomposition \label{TED_framework}}

Many-fermion methods, especially those in quantum chemistry,
 are often arranged as a systematic hierarchy of increasing cost and accuracy.
While it is possible to construct a TED for a flexible choice
 of operator basis limited only by Eq. (\ref{V.sparse}),
 this section considers only a specific hierarchy of methods.
The methods are referred to as TED$r$ for an integer $r$
 and defined by the model operator subspace
\begin{equation}
 \{ \hat{C}(\mathbf{f},\mathbf{g}) : \|\mathbf{f}\cup\mathbf{g}\| \le r \}.
\end{equation}
TED$r$ is exact for $r = n$ and accuracy should systematically improve with increasing $r$.
Computational scaling of the methods depend on $r$
 and the total number of fermion degrees of freedom $n$.
The memory required to store each operator scales as $O(n^r)$.
Commutation of operators is the most computationally expensive step of the TED
 and scales as $O(n^{\lfloor1.5r\rfloor})$ operations.
$O(n^r)$ memory and $O(n^{\lfloor1.5r\rfloor})$ operations are taken to be
 the computational budget of TED$r$ for calculating physical properties
 following the diagonalization of a Hamiltonian.

A Hamiltonian diagonalized by TED$r$ as in Eq. (\ref{approx_diag}) has the form
\begin{equation}
 \hat{D} = \sum_{\|\mathbf{f}\| \le r} d(\mathbf{f}) \hat{C}(\mathbf{f},\mathbf{f}).
\end{equation}
An eigenvalue $E(\mathbf{z})$ can be calculated with this formula
 by replacing $\hat{n}_i$ with $z_i$.
This calculation scales as $O(n^r)$ operations.
Only $O(n^{\lfloor1.5r\rfloor/r})$ eigenvalues can be calculated
 this way before the computational budget is exhausted.
A specialized alternative is to transform $\hat{D}$
 into the $\hat{E}_V^{\mathbf{z}}(\mathbf{f},\mathbf{f})$ operator basis,
\begin{equation} \label{excitations}
 \hat{D} = \sum_{\|\mathbf{f}\| \le r} E(\mathbf{f}\oplus\mathbf{z}) \hat{E}_V^{\mathbf{z}}(\mathbf{f},\mathbf{f}).
\end{equation}
This simultaneously calculates $O(n^r)$ eigenvalues
 corresponding to few-eigenfermion excitations
 from a reference eigenstate $|\Psi(\mathbf{z})\rangle$.
Each of these transformations costs $O(n^r)$ operations, which increases
 the total number of computable eigenvalues to $O(n^{\lfloor1.5r\rfloor})$.

It is in the study of energetics that the TED can be considered a quantum-to-classical mapping.
All energies come from $E(\mathbf{f})$ in Eq. (\ref{eigenvalue}), which for TED$r$
 has a form that resembles the total energy cluster expansions used in the study of alloys \cite{Sanchez.cluster}.
The complexity of finding the ground state is dramatically reduced from the initial Hamiltonian,
\begin{equation}
 \mathrm{from} \ \ \min_{|\Psi\rangle} \frac{\langle\Psi|\hat{H}|\Psi\rangle}{ \langle\Psi|\Psi\rangle} \ \ \mathrm{to} \ \ \min_{\mathbf{f}} E(\mathbf{f}).
\end{equation}
A minimization over $2^n$ complex numbers is reduced to $n$ binary choices.
Complexity theory still classifies both problems as hard,
 QMA-complete for the quantum problem \cite{Kempe.Qcomplexity} and 
 NP-complete for the classically-mapped problem \cite{Barahona.hardmin}.
Practically, many physical systems of interest are unfrustrated and
 will result in easy instances of minimization over $E(\mathbf{f})$.
Even in hard cases, simple heuristics can give good results.
Eq. (\ref{excitations}) can be calculated for random sets of $\mathbf{z}$
 and energy-lowering few-fermion excitations can be successively applied.
In an easy problem, most or all initial configurations will relax to the ground state.
A hard ``glassy'' problem will have many distinct local minima in configuration space.
In ``exotic'' systems where the low-energy excitations are not few-eigenfermion excitations,
 this procedure might help to map out the energy landscape.

To calculate eigenstate matrix elements of an operator $\hat{X}$,
 a truncated unitary transformation is performed.
This exactly calculates $\hat{U}^\dag \hat{X}' \hat{U}$ for a perturbed operator
 $\hat{X}' = \hat{X} - \hat{U} \hat{X}_R \hat{U}^\dag$ as in Eq. (\ref{error_form}).
The transformation costs $O(n^{\lfloor1.5r\rfloor})$ operations,
 which exceeds the computational budget if more than $O(1)$ operators are calculated in this manner.
Calculations of this nature are able to produce a large amount of spectral information
 for a small number of operators.
This might be useful for categorizing eigenstates and transitions
 based on a small number of important observables.
By rewriting the transformed operator in the $\hat{E}_V^{\mathbf{z}}(\mathbf{f},\mathbf{g})$ operator basis,
\begin{equation}
 \hat{U}^\dag \hat{X}' \hat{U} = \sum_{\|\mathbf{f}\cup\mathbf{g}\| \le r} \langle \Psi(\mathbf{f}\oplus\mathbf{z}) | \hat{X}' |\Psi(\mathbf{g}\oplus\mathbf{z})\rangle \hat{E}_V^{\mathbf{z}}(\mathbf{f},\mathbf{g}),
\end{equation}
 a set of approximate matrix elements close to a reference
 eigenstate $|\Psi(\mathbf{z})\rangle$ can be efficiently computed.

The calculation of reduced density matrices is an established
 application of truncated unitary transformations \cite{Mazziotti.ACSE}.
They can be calculated as a subset of $\langle \Psi(\mathbf{z})|\hat{C}(\mathbf{f},\mathbf{g})|\Psi(\mathbf{z})\rangle$
 for a chosen eigenstate and all elements of the model subspace.
To close this calculation, the infinitesimal transformation of a matrix element
 must be related back to the untransformed matrix elements.
The key step is performing the transformation backwards
 by defining a new unitary operator $\hat{V}(\lambda)$ that evolves as
\begin{equation}
  \frac{d}{d\lambda} \hat{V}(\lambda) = -\hat{A}(\lambda_F - \lambda)\hat{V}(\lambda),
\end{equation}
 with initial condition $\hat{V}(0) = \hat{I}$.
The entire transformation that defines $\hat{U}$ must occur in the interval $[0,\lambda_F]$,
 resulting in $\hat{V}(\lambda_F) = \hat{U}$.
Using the same truncation as in the transformation of operators,
 the truncated transformation of matrix elements is defined as
\begin{equation}
  \frac{d}{d\lambda} \langle \hat{C}(\mathbf{f},\mathbf{g}) \rangle (\lambda) = \langle \widehat{\mathcal{P}}_M [\hat{A}(\lambda_F - \lambda),\hat{C}(\mathbf{f},\mathbf{g})] \rangle (\lambda),
\end{equation}
 with $\langle \hat{C}(\mathbf{f},\mathbf{g}) \rangle (\lambda) \approx \langle \mathbf{z}| \hat{V}^\dag(\lambda) \hat{C}(\mathbf{f},\mathbf{g}) \hat{V}(\lambda) |\mathbf{z}\rangle$.
For a piecewise constant generator, the transformation can be evaluated as a sequence of
 superoperator exponentials,
\begin{equation}
  \langle \hat{C}(\mathbf{f},\mathbf{g}) \rangle_i = \langle \exp(\widehat{\mathcal{A}}_{m-i+1}) \hat{C}(\mathbf{f},\mathbf{g}) \rangle_{i-1},
\end{equation}
 where $m$ is the number of generator segments.
As with operator transformations, the cost of this calculation scales as $O(n^{\lfloor1.5r\rfloor})$ operations.

\begin{acknowledgments}
I thank Jeff Hammond for useful discussions in the early stages of this work.
I thank Jay Deep Sau for valuable criticism on early drafts of the paper.
This work was supported by the National Science Foundation under the grant DMR-09-41645.
\end{acknowledgments}

\bibliography{$HOME/work/my.bib}

\begin{thebibliography}{10}%
\makeatletter
\providecommand \@ifxundefined [1]{%
 \ifx #1\undefined \expandafter \@firstoftwo
 \else \expandafter \@secondoftwo
\fi
}%
\providecommand \@ifnum [1]{%
 \ifnum #1\expandafter \@firstoftwo
 \else \expandafter \@secondoftwo
\fi
}%
\providecommand \enquote [1]{``#1''}%
\providecommand \bibnamefont  [1]{#1}%
\providecommand \bibfnamefont [1]{#1}%
\providecommand \citenamefont [1]{#1}%
\providecommand\href[0]{\@sanitize\@href}%
\providecommand\@href[1]{\endgroup\@@startlink{#1}\endgroup\@@href}%
\providecommand\@@href[1]{#1\@@endlink}%
\providecommand \@sanitize [0]{\begingroup\catcode`\&12\catcode`\#12\relax}%
\@ifxundefined \pdfoutput {\@firstoftwo}{%
 \@ifnum{\z@=\pdfoutput}{\@firstoftwo}{\@secondoftwo}%
}{%
 \providecommand\@@startlink[1]{\leavevmode\special{html:<a href="#1">}}%
 \providecommand\@@endlink[0]{\special{html:</a>}}%
}{%
 \providecommand\@@startlink[1]{%
  \leavevmode
  \pdfstartlink
   attr{/Border[0 0 1 ]/H/I/C[0 1 1]}%
   user{/Subtype/Link/A<</Type/Action/S/URI/URI(#1)>>}%
  \relax
 }%
 \providecommand\@@endlink[0]{\pdfendlink}%
}%
\providecommand \url  [0]{\begingroup\@sanitize \@url }%
\providecommand \@url [1]{\endgroup\@href {#1}{\urlprefix}}%
\providecommand \urlprefix [0]{URL }%
\providecommand \Eprint[0]{\href }%
\@ifxundefined \urlstyle {%
  \providecommand \doi [1]{doi:\discretionary{}{}{}#1}%
}{%
  \providecommand \doi [0]{doi:\discretionary{}{}{}\begingroup
  \urlstyle{rm}\Url }%
}%
\providecommand \doibase [0]{http://dx.doi.org/}%
\providecommand \Doi[1]{\href{\doibase#1}}%
\providecommand \bibAnnote [3]{%
  \BibitemShut{#1}%
  \begin{quotation}\noindent
    \textsc{Key:}\ #2\\\textsc{Annotation:}\ #3%
  \end{quotation}%
}%
\providecommand \bibAnnoteFile [2]{%
  \IfFileExists{#2}{\bibAnnote {#1} {#2} {\input{#2}}}{}%
}%
\providecommand \typeout [0]{\immediate \write \m@ne }%
\providecommand \selectlanguage [0]{\@gobble}%
\providecommand \bibinfo [0]{\@secondoftwo}%
\providecommand \bibfield [0]{\@secondoftwo}%
\providecommand \translation [1]{[#1]}%
\providecommand \BibitemOpen[0]{}%
\providecommand \bibitemStop [0]{}%
\providecommand \bibitemNoStop [0]{.\EOS\space}%
\providecommand \EOS [0]{\spacefactor3000\relax}%
\providecommand \BibitemShut [1]{\csname bibitem#1\endcsname}%
\bibitem{Olsen.FCI}%
  \BibitemOpen
  \bibfield{author}{%
  \bibinfo {author} {\bibfnamefont{J.}~\bibnamefont{Olsen}}, \bibinfo {author}
  {\bibfnamefont{P.}~\bibnamefont{J{\o}rgensen}},\ and\ \bibinfo {author}
  {\bibfnamefont{J.}~\bibnamefont{Simons}},\ }%
  \bibfield{journal}{%
  \bibinfo {journal} {Chem. Phys. Lett.}\ }%
  \textbf{\bibinfo {volume} {169}},\ \bibinfo {pages} {463} (\bibinfo {year}
  {1990})%
  \bibAnnoteFile{NoStop}{Olsen.FCI}%
\bibitem{Onida.TDDFT.GWBSE}%
  \BibitemOpen
  \bibfield{author}{%
  \bibinfo {author} {\bibfnamefont{G.}~\bibnamefont{Onida}}, \bibinfo {author}
  {\bibfnamefont{L.}~\bibnamefont{Reining}},\ and\ \bibinfo {author}
  {\bibfnamefont{A.}~\bibnamefont{Rubio}},\ }%
  \bibfield{journal}{%
  \bibinfo {journal} {Rev. Mod. Phys.}\ }%
  \textbf{\bibinfo {volume} {74}},\ \bibinfo {pages} {601} (\bibinfo {year}
  {2002})%
  \bibAnnoteFile{NoStop}{Onida.TDDFT.GWBSE}%
\bibitem{Foulkes.QMC}%
  \BibitemOpen
  \bibfield{author}{%
  \bibinfo {author} {\bibfnamefont{W.~M.~C.}\ \bibnamefont{Foulkes}}, \bibinfo
  {author} {\bibfnamefont{L.}~\bibnamefont{Mitas}}, \bibinfo {author}
  {\bibfnamefont{R.~J.}\ \bibnamefont{Needs}},\ and\ \bibinfo {author}
  {\bibfnamefont{G.}~\bibnamefont{Rajagopal}},\ }%
  \bibfield{journal}{%
  \bibinfo {journal} {Rev. Mod. Phys.}\ }%
  \textbf{\bibinfo {volume} {73}},\ \bibinfo {pages} {33} (\bibinfo {year}
  {2001})%
  \bibAnnoteFile{NoStop}{Foulkes.QMC}%
\bibitem{Schollwock.DMRG}%
  \BibitemOpen
  \bibfield{author}{%
  \bibinfo {author} {\bibfnamefont{U.}~\bibnamefont{Schollw{\"o}ck}},\ }%
  \bibfield{journal}{%
  \bibinfo {journal} {Rev. Mod. Phys.}\ }%
  \textbf{\bibinfo {volume} {77}},\ \bibinfo {pages} {259} (\bibinfo {year}
  {2005})%
  \bibAnnoteFile{NoStop}{Schollwock.DMRG}%
\bibitem{Georges.DMFT}%
  \BibitemOpen
  \bibfield{author}{%
  \bibinfo {author} {\bibfnamefont{A.}~\bibnamefont{Georges}}, \bibinfo
  {author} {\bibfnamefont{G.}~\bibnamefont{Kotliar}}, \bibinfo {author}
  {\bibfnamefont{W.}~\bibnamefont{Krauth}},\ and\ \bibinfo {author}
  {\bibfnamefont{M.~J.}\ \bibnamefont{Rozenberg}},\ }%
  \bibfield{journal}{%
  \bibinfo {journal} {Rev. Mod. Phys.}\ }%
  \textbf{\bibinfo {volume} {68}},\ \bibinfo {pages} {13} (\bibinfo {year}
  {1996})%
  \bibAnnoteFile{NoStop}{Georges.DMFT}%
\bibitem{Wegner.flow}%
  \BibitemOpen
  \bibfield{author}{%
  \bibinfo {author} {\bibfnamefont{F.}~\bibnamefont{Wegner}},\ }%
  \bibfield{journal}{%
  \bibinfo {journal} {J. Phys. A}\ }%
  \textbf{\bibinfo {volume} {39}},\ \bibinfo {pages} {8221} (\bibinfo {year}
  {2006})%
  \bibAnnoteFile{NoStop}{Wegner.flow}%
\bibitem{White.canonical.diagonalization}%
  \BibitemOpen
  \bibfield{author}{%
  \bibinfo {author} {\bibfnamefont{S.~R.}\ \bibnamefont{White}},\ }%
  \bibfield{journal}{%
  \bibinfo {journal} {J. Chem. Phys.}\ }%
  \textbf{\bibinfo {volume} {117}},\ \bibinfo {pages} {7472} (\bibinfo {year}
  {2002})%
  \bibAnnoteFile{NoStop}{White.canonical.diagonalization}%
\bibitem{Yanai.canonical.transformation}%
  \BibitemOpen
  \bibfield{author}{%
  \bibinfo {author} {\bibfnamefont{T.}~\bibnamefont{Yanai}}\ and\ \bibinfo
  {author} {\bibfnamefont{G.~K.-L.}\ \bibnamefont{Chan}},\ }%
  \bibfield{journal}{%
  \bibinfo {journal} {J. Chem. Phys.}\ }%
  \textbf{\bibinfo {volume} {127}},\ \bibinfo {eid} {104107} (\bibinfo {year}
  {2007})%
  \bibAnnoteFile{NoStop}{Yanai.canonical.transformation}%
\bibitem{Wegner.oldflow}%
  \BibitemOpen
  \bibfield{author}{%
  \bibinfo {author} {\bibfnamefont{F.}~\bibnamefont{Wegner}},\ }%
  \bibfield{journal}{%
  \bibinfo {journal} {Ann. Physik}\ }%
  \textbf{\bibinfo {volume} {3}},\ \bibinfo {pages} {77} (\bibinfo {year}
  {1994})%
  \bibAnnoteFile{NoStop}{Wegner.oldflow}%
\bibitem{Taube.UCC}%
  \BibitemOpen
  \bibfield{author}{%
  \bibinfo {author} {\bibfnamefont{A.~G.}\ \bibnamefont{Taube}}\ and\ \bibinfo
  {author} {\bibfnamefont{R.~J.}\ \bibnamefont{Bartlett}},\ }%
  \bibfield{journal}{%
  \bibinfo {journal} {Int. J. Quantum Chem.}\ }%
  \textbf{\bibinfo {volume} {106}},\ \bibinfo {pages} {3393} (\bibinfo {year}
  {2006})%
  \bibAnnoteFile{NoStop}{Taube.UCC}%
\bibitem{Hubsch.projector.renormalization}%
  \BibitemOpen
  \bibfield{author}{%
  \bibinfo {author} {\bibfnamefont{A.}~\bibnamefont{H{\"u}bsch}}, \bibinfo
  {author} {\bibfnamefont{S.}~\bibnamefont{Sykora}},\ and\ \bibinfo {author}
  {\bibfnamefont{K.~W.}\ \bibnamefont{Becker}}\ }%
  \Eprint{http://arxiv.org/abs/0809.3360}{arXiv:0809.3360}%
  \bibAnnoteFile{NoStop}{Hubsch.projector.renormalization}%
\bibitem{Mazziotti.ACSE}%
  \BibitemOpen
  \bibfield{author}{%
  \bibinfo {author} {\bibfnamefont{D.~A.}\ \bibnamefont{Mazziotti}},\ }%
  \bibfield{journal}{%
  \bibinfo {journal} {Phys. Rev. Lett.}\ }%
  \textbf{\bibinfo {volume} {97}},\ \bibinfo {eid} {143002} (\bibinfo {year}
  {2006})%
  \bibAnnoteFile{NoStop}{Mazziotti.ACSE}%
\bibitem{Wildberger.lie_diag}%
  \BibitemOpen
  \bibfield{author}{%
  \bibinfo {author} {\bibfnamefont{N.~J.}\ \bibnamefont{Wildberger}},\ }%
  \bibfield{journal}{%
  \bibinfo {journal} {P. Am. Math. Soc.}\ }%
  \textbf{\bibinfo {volume} {119}},\ \bibinfo {pages} {649} (\bibinfo {year}
  {1993})%
  \bibAnnoteFile{NoStop}{Wildberger.lie_diag}%
\bibitem{Bartlett.CCreview}%
  \BibitemOpen
  \bibfield{author}{%
  \bibinfo {author} {\bibfnamefont{R.~J.}\ \bibnamefont{Bartlett}}\ and\
  \bibinfo {author} {\bibfnamefont{M.}~\bibnamefont{Musia{\l}}},\ }%
  \bibfield{journal}{%
  \bibinfo {journal} {Rev. Mod. Phys.}\ }%
  \textbf{\bibinfo {volume} {79}},\ \bibinfo {eid} {291} (\bibinfo {year}
  {2007})%
  \bibAnnoteFile{NoStop}{Bartlett.CCreview}%
\bibitem{Parlett.max_ev}%
  \BibitemOpen
  \bibfield{author}{%
  \bibinfo {author} {\bibfnamefont{B.~N.}\ \bibnamefont{Parlett}}, \bibinfo
  {author} {\bibfnamefont{H.}~\bibnamefont{Simon}},\ and\ \bibinfo {author}
  {\bibfnamefont{L.~M.}\ \bibnamefont{Stringer}},\ }%
  \bibfield{journal}{%
  \bibinfo {journal} {Math. Comput.}\ }%
  \textbf{\bibinfo {volume} {38}},\ \bibinfo {pages} {153} (\bibinfo {year}
  {1982})%
  \bibAnnoteFile{NoStop}{Parlett.max_ev}%
\bibitem{Sanchez.cluster}%
  \BibitemOpen
  \bibfield{author}{%
  \bibinfo {author} {\bibfnamefont{J.~M.}\ \bibnamefont{Sanchez}}, \bibinfo
  {author} {\bibfnamefont{F.}~\bibnamefont{Ducastelle}},\ and\ \bibinfo
  {author} {\bibfnamefont{D.}~\bibnamefont{Gratias}},\ }%
  \bibfield{journal}{%
  \bibinfo {journal} {Physica A}\ }%
  \textbf{\bibinfo {volume} {128}},\ \bibinfo {pages} {334} (\bibinfo {year}
  {1984})%
  \bibAnnoteFile{NoStop}{Sanchez.cluster}%
\bibitem{Kempe.Qcomplexity}%
  \BibitemOpen
  \bibfield{author}{%
  \bibinfo {author} {\bibfnamefont{J.}~\bibnamefont{Kempe}}, \bibinfo {author}
  {\bibfnamefont{A.}~\bibnamefont{Kitaev}},\ and\ \bibinfo {author}
  {\bibfnamefont{O.}~\bibnamefont{Regev}},\ }%
  \bibfield{journal}{%
  \bibinfo {journal} {SIAM J. Comput.}\ }%
  \textbf{\bibinfo {volume} {35}},\ \bibinfo {pages} {1070} (\bibinfo {year}
  {2006})%
  \bibAnnoteFile{NoStop}{Kempe.Qcomplexity}%
\bibitem{Barahona.hardmin}%
  \BibitemOpen
  \bibfield{author}{%
  \bibinfo {author} {\bibfnamefont{F.}~\bibnamefont{Barahona}},\ }%
  \bibfield{journal}{%
  \bibinfo {journal} {J. Phys. A}\ }%
  \textbf{\bibinfo {volume} {15}},\ \bibinfo {pages} {3241} (\bibinfo {year}
  {1982})%
  \bibAnnoteFile{NoStop}{Barahona.hardmin}%
\end{thebibliography}%

\end{document}